\documentclass[12pt,onecolumn, peerreview,draftclsnofoot]{IEEEtran} 
\ifCLASSINFOpdf
   \usepackage[pdftex]{graphicx}
\else
   \usepackage[dvips]{graphicx}  
\fi
\graphicspath{{figures/}}
\usepackage[cmex10]{amsmath}
\usepackage{amsthm}
\usepackage{array}
\usepackage{eqparbox}
\usepackage{txfonts}
\usepackage{verbatim, amssymb}
\usepackage{mathrsfs}
\usepackage{bm}
\usepackage{lipsum}
\usepackage{cite}
\usepackage{algpseudocode}
\usepackage{algorithmicx}
\usepackage{algpseudocode}
\usepackage{algorithm}
\usepackage{textcomp}
\usepackage{pst-circ}
\usepackage{multibib}
\usepackage{mathrsfs}
\usepackage{multicol}
\usepackage{mathtools, cuted}
\usepackage{notoccite}

\begin{document}
 
\title{Underpinnings of User-Channel Allocation in Non-Orthogonal Multiple Access for 5G} 

\author{Shabnam~Sodagari,~\IEEEmembership{Senior Member,~IEEE} 

}

 \maketitle
\begin{abstract}
Non-orthogonal multiple access (NOMA) is a part of 5th generation (5G) communication systems. 
This article presents the underpinnings and underlying structures of the problem of NOMA user-channel allocation. Unlike the heuristics for NOMA user-channel allocation, the presented results are guaranteed to converge to a solution. In addition, the solutions are stable. Generally, the results apply to any NOMA system. Unlike the orthogonal frequency division multiple access (OFDMA) resource allocation problem, the core matching is not the solution to NOMA resource allocation. The conditions under which the fix-point NOMA resource allocation is guaranteed to be stable from the viewpoint of both the base station and the NOMA users are described. In addition, relationships of NOMA user-channel resource allocation to game models and subgame perfect Nash equilibria are elucidated. 

\end{abstract}

\IEEEpeerreviewmaketitle
\section{Introduction}
Vehicular communication is an important element of smart cities. It is an application of the Internet of Things (IoT) that is anticipated to be part of 5G technology for smart transport. One advantage of cellular vehicle to everything (V2X) communications is deploying the Long-term evolution (LTE) infrastructure that is already in place in ubiquitous areas.

Cellular V2X (C-V2X) is preferred to IEEE 802.11p according to 3GPP Release 14. The standard IEEE 802.11p is a WiFi-like standard for vehicles. It belongs to the family of IEEE 802.11 standard series. The US Department of Transportation (USDoT) had previously proposed DSRC (Dedicated Short-Range Communications), based on IEEE 802.11p. However, C-V2X offers wider coverage on both freeways and intra-city roads, lower delays, and higher reliability~\cite{3ofkhor,khorya}. Therefore, a paradigm shift in vehicular communications is in progress from WiFi-like technologies toward LTE-like solutions, since the cellular infrastructure is already available.

The onboard safety features of vehicles can benefit from C-V2X communications, e.g., for blind spots, road hazard information dissemination, left turn assist (LTA), active rollover protection (ARP), etc. 

On the other side, the 5G cellular systems are adopting non-orthogonal multiple access (NOMA) over the conventionally used orthogonal frequency division multiple access (OFDMA)~\cite{shober}. The same principle applies to C-V2X with the ever-increasing number of vehicles that are vying to connect to the network and to other vehicles for different services. Orthogonal multiple access methods fall short in terms of accommodating the large number of users envisioned for 5G. Consequently, NOMA is a solution that can provide a lower delay, prevent congestion, and meet network traffic demands~\cite{cvxnoma}.
Another benefit of NOMA LTE systems is higher spectral efficiency compared to orthogonal multiple access.

Two major premises of NOMA are power domain multiplexing and code domain
multiplexing. In power domain NOMA, different users are allocated different power
levels according to their channel state information. The difference between power levels and channel gains allow successive interference cancellation (SIC) at the decoder to cancel interference. In code domain multiplexing NOMA, different codes are assigned to different users to enable multiuser communications. In this aspect, code domain NOMA is similar to code division multiple access (CDMA)~\cite{NOMAcomm}.  

This article presents a stable user-channel allocation scheme for NOMA LTE cellular V2X systems. The \textit{main contributions} of this article is presenting a user-channel allocation NOMA scheme that always converges to a solution. In addition, the scheme always leads to a stable and optimal solution in which no channel could be allocated to a better user and no user could be assigned to transmit over a better channel if they deviate from this scheme. 

The rest of the paper is organized as follows. Section~\ref{sec:model} presents results on the novel user-channel allocation scheme in NOMA cellular V2X communications. 
Section~\ref{sec:conclus} concludes the paper. Table~\ref{table:not} contains  abbreviations and the notation used throughout the paper. 

\begin{table*}
\fontsize{11}{11}
	\centering
\caption{Abbreviations}
\begin{tabular}{l|l}
  \hline
  $\mathscr{V}$ & The set of vehicles\\\hline
	$\mathscr{C}$ & The set of channels \\\hline 
   $\mathfrak{P}(\mathscr{V})$ & The set of all subsets of $\mathscr{V}$ \\\hline
 	$\mathfrak{P}(\mathscr{C})$ &The set of all subsets of $\mathscr{C}$ \\\hline
	$P(v)$ & The preference ordering of vehicle $v$ defined over $\mathfrak{P}(\mathscr{C})$ \\\hline
$P(c)$ & The preference ordering of channel $c$ defined over $\mathfrak{P}(\mathscr{V})$ \\\hline
	$M$ & A NOMA user-channel allocation \\\hline
	$M^{\prime} $ & A blocking user-channel allocation to $M$ \\\hline
	$U(c,M)$ & The set of vehicles that are willing to transmit on channel $c$ even after dropping some of the channels, assigned by $M(v)$ \\\hline
	$V(v,M)$ & The set of channels that the base station is willing to allocate to $v$ even after dropping some of the vehicles, assigned by $M(c)$\\\hline
   $TM(c)$ & channel $c$'s optimal set of vehicles among those willing to transmit on $c$ \\\hline
	$TM(v)$ & The set of channels preferred by $v$ among the channels that the base station is willing to allocate to $v$\\\hline
\end{tabular}
\vspace{-.1in}
\label{table:not}
\end{table*} 

\section{System Model}
\label{sec:model}

In our scheme, NOMA allows each channel to be allocated to more than one vehicle and each vehicle can use more than one channel. This article has further aimed to study the underlying structures of the problem of user-channel allocation in NOMA.

We denote the set of users (here, vehicles) by $\mathscr{V}$ and the set of channels by $\mathscr{C}$. We denote the power set, i.e., the set of all subsets of $\mathscr{V}$ and $\mathscr{C}$ by $\mathfrak{P}(\mathscr{V})$ and $\mathfrak{P}(\mathscr{C})$, respectively. Using NOMA, the base station allocates every channel to a subset of vehicles, i.e., an element of $\mathfrak{P}(\mathscr{V})$. In addition, every vehicle is allocated to a subset of channels, taken from $\mathfrak{P}(\mathscr{C})$. The base station has geolocation information of the users. The allocation can be based on various criteria, such as geolocation data, maximum data rate, minimum error rate, etc.  
To this end, we note that every vehicle $v\in \mathscr{V}$ has a preference $P(v)$ over $\mathfrak{P}(\mathscr{C})$, which means preferring a subset of channels over another subset of channels. Similarly, every channel $c\in \mathscr{C}$ has a preference $P(c)$ that prefers a subset of vehicles to another subset of vehicles, e.g., according to the ergodic capacity, geolocation, etc. For any vehicle $v\in \mathscr{V}$, the set of allocations is a subset of $\mathscr{C}$, and for any channel $c\in \mathscr{C}$, the set of allocations is a subset of $\mathscr{V}$.

We denote the NOMA user-channel allocation by the matching $M$. We note that a NOMA user-channel allocation is called \textbf{individually rational} if neither the base station nor the user prefer to cancel some of the channel allocations \cite{contracts}. Not all matches can be suitable user-channel allocations in NOMA. A good NOMA user-channel allocation should not be blocked. Blocking can involve the following case called \textbf{pairwise blocking}: There is a user-channel pair in which the user in this pair would like to add a new channel or drop a previously allocated channel, or the base station would like to cancel a channel allocated to the user and allocate this channel to another user. A lack of pairwise blocking leads to \textbf{pairwise stability}. NOMA user-channel matching is \textbf{pairwise stable} if it is individually rational and there exists no user-channel pair that can pairwise block the matching. 

In some cases, if the NOMA user-channel allocation is not effective, more than just one user-channel pair are involved in the blocking. This leads us to consider 
\textbf{setwise blocking}, wherein a set of channels and users would like to implement a new set of allocations among themselves while possibly cancelling other allocations. Setwise blockings also require that the blocking allocation is individually rational, i.e., none of the members of the blocking coalition has an incentive to cancel the allocations unilaterally~\cite{contracts}. 

For individually rational allocations, an example of a setwise block is a coalition block. The coalition block comprises a group of users and channels such that the members of the coalition get better connection services by adding some new matches $M^{\prime}$ within the coalition while possibly canceling the current NOMA resource allocations, as determined by $M$. As such, the vehicles and channels outside of the blocking coalition are not assigned new matches, but possibly some of their matching allocations are canceled by members of the blocking coalition. This shows that if the NOMA resource allocation is not optimal, it will be fragile in that the members of the blocking coalition prefer rearrangements to obtain new resources that were not provided to them by the scheduling algorithm. The matching $M^{\prime}$ is called a blocking allocation for the matching $M$, since $M^{\prime}$ setwise blocks $M$. For all the members of the blocking coalition, the new resource allocation $M^{\prime}$  is preferred to the old set of matching $M$. The new matching $M^{\prime}$ is implemented among the members of the blocking coalition only. The user-channel pairs outside of the blocking coalition preserve their old allocations, as determined by $M$. The user-channel pairs inside the blocking coalition all receive a better and individually rational set of allocations. In fact, this is the main incentive for the formation of the blocking coalition among a subset of users and a subset of channels.

After defining the individual rationality and blockings for user-channel allocations, we are now ready to define a stable NOMA user-channel allocation, which is a robust and efficient allocation. A \textbf{stable NOMA user-channel allocation} is individually rational (as defined above) and does not involve any blocking. It is important for the NOMA resource allocation algorithm to be stable. In addition, being individually rational is a necessary condition for the stability of NOMA resource allocation. 

The NOMA user-channel assignment consists of allocating more than one user to a channel while any user can transmit over several channels, as long as the user's transceiver has the hardware/software radio capabilities for transmission over multiple channels. Therefore, this problem can be cast to  many-to-many matching. It is shown that in many-to-many matching, the core is not the same as the set of (setwise) stable matchings, unlike one-to-one matching. This is a consequence of the fact that in the core allocation, the users have an incentive to unilaterally cancel their allocations~\cite{contracts}. Therefore, the core is not a solution to NOMA user-channel resource allocation. Accordingly, to derive a stable NOMA resource allocation, we need to eliminate blockings as defined before. More specifically, we need to find matchings that are both pairwise-stable and setwise-stable, i.e., that have neither pairwise blocks nor setwise blocks. In this regard, we consider the concept of substitutability and then relate it to finding the solution of stable NOMA resource allocation. 

In simple words, the notion of \textbf{substitutability} implies that if the base station (on behalf of a particular channel) does not choose a vehicle from some set of vehicles, then that vehicle is still not chosen by the base station from a larger set of vehicles. Similarly, if the preference $P(v)$ of a vehicle $v\in \mathscr{V}$ does not choose a channel from some set of channels, then that channel will not be chosen from a larger set of channels~\cite{milgrom}. 

In other words, the preference $P(v)$ of a vehicle $v\in \mathscr{V}$ (or $P(c)$ of a channel $c \in \mathscr{C}$) is called substitutable if it meets the following condition: when a vehicle $v$ (or channel $c$) selects a channel (vehicle) among some set of channels (vehicles), then the same channel (vehicle) will be chosen by $v$ (or $c$) from a smaller set of channels (vehicles). 

In \textbf{strong substitutable} preference conditions~\cite{oveido} if a vehicle $v$ prefers a channel $c$ among some set of channels, then that vehicle selects the same channel $c$ from a worse set of channels that includes it. In other terms, if the vehicle $v$ does not select a particular channel from some set of channels, then, that channel will not be selected by $v$ from a set of better channels for $v$, according to the preference profile $P(v)$. 

Strong substitutability is a sufficient condition for substitutability. The reason we are interested in strong substitutability is due to its relation to stability of NOMA resource allocations. If preferences are strongly substitutable, then the set of pairwise stable allocations is equal to the set of setwise stable resource allocations~\cite{oveido}.

\subsection{Differences with OFDMA}

In OFDMA, only one user is allocated to every channel, whereas in NOMA multiple users are allocated to one channel. In addition, each user can transmit over two or more channels, as long as the user has the necessary configurations. Therefore, in OFDMA, core matching is a solution for channel allocation, since it is individually rational and pairwise stable. However, unlike OFDMA, in NOMA wherein multiple users transmit over a channel and multiple channels are allocated to a user (many-to-many matching), core matching is not a solution, since it may not be individually rational. In other words, the core channel allocation may provide a solution in which a user $v_1 \in \mathscr{V}$ may get better quality of service (QoS) over another channel $c_i \in \mathscr{C}$ compared to its current assigned channel $c_j \in \mathscr{C}$ (by core matching) and at the same time the allocation of $c_i$ to another user  $v_2 \in \mathscr{V}$ would give better performance than allocating it to user $v_1$. This violation of pairwise stability shows us that unlike OFDMA, in NOMA, core matching is not a stable solution to the channel allocation problem. 

The solutions for the NOMA channel allocation problem can be divided into the following categories:
\begin{enumerate}
\item setwise stable channel/user allocations, i.e., individually rational channel/user allocation that cannot be blocked by a coalition~\cite{roth1984,sotomayor1999};
\item individually rational core: the set of individually rational matchings that cannot be blocked~\cite{sotomayor1999};
\item a user-channel allocation that meets pairwise stability property, i.e., without any pairwise blocking; and
\item the set of fix-points of a function~\cite{oveido}: In the matching that is described in this article, each user $v$ (or the base station on behalf of a channel $c$) is selecting the most preferred channels (or users), among the set of potential channels (or users) that also prefer to be matched to $v$ (or $c$).

\end{enumerate}

\subsection{NOMA User-Channel Allocation as a Bargaining Game}

Another interpretation of NOMA user-channel resource allocation is a non-cooperative bargaining game~\cite{oveido}. The subgame perfect Nash equilibrium is used for sequential games. It implies that not only the outcomes of the game need to be in a Nash equilibrium, but in every subgame the strategy profile needs to be in a Nash
equilibrium~\cite{subgame}. A subgame is the piece of a game that
is played starting at any point at which the complete history of the game played so far is common knowledge~\cite{gibb}.

\begin{figure}[htb!]
\centering
\includegraphics[width=3.5in]{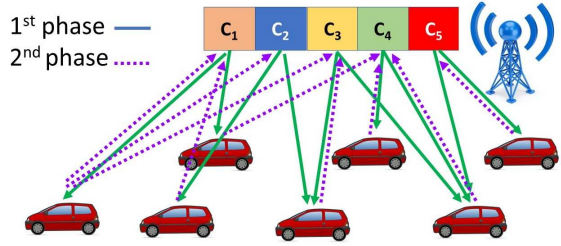}
\caption{The subgame played by the set of users and the set of channels related to setwise-stable NOMA resource allocation} 
\label{fig:spne}
\end{figure}

The main reason behind this model as shown in Fig.~\ref{fig:spne} is that the set of subgame-perfect Nash equilibrium (SPNE) outcomes of the game has the setwise stability property considering a complete-information setting. As mentioned before, stability is an important factor for NOMA resource allocation. 

The two types of players of this sequential extensive-form game are the channels (represented by the cellular base station) and the NOMA users (here the vehicles). The strategy of each player involves the selection of a subset of the other type of players. In the first round, the base station simultaneously sends messages (on behalf of every channel $c$) to a subset of vehicles. This strategy can be based on a criterion, such as the geolocation information of the users, available at the base station. After the vehicles receive the base station message, each vehicle takes part in the second round of the subgame by selecting a subset of channels, based on desired QoS requirements. Finally, a matching $M$ is achieved by matching a vehicle $v$ to a channel $c$ if and only if $v$ is in the subset for $c$ in the first round of the subgame and $c$ is in the subset for $v$ in the second round. 

\subsection{Stable NOMA User-Channel Allocation}

The details of the stable NOMA user-channel allocation scheme are outlined in Algorithm~\ref{algmatch}. 

\begin{algorithm}
\caption{Stable NOMA User-Channel Resource Allocation}\label{algmatch}
\begin{algorithmic}[1]
\State \textbf{Input:} $P(v), P(c), \forall v\in \mathscr{V}, \forall c \in \mathscr{C}$\Comment {Preference ordering of channels and vehicles based on base station geolocation data and hardware/software radios of the vehicles.} 
\State {\bf Step 1:} Match $M(c)= \emptyset,~M(v) =\mathscr{C}$ \Comment{Initialization}
\State {\bf Step 2:} Find $U(c,M)$ \Comment{The set of vehicles that are willing to transmit on channel $c$, possibly after dropping some of the channels, assigned by $M(v)$. }
\State �\indent Find $V(v,M)$ \Comment{The set of channels that the base station is willing to allocate to $v$, possibly after dropping some of the vehicles, assigned by $M(c)$.} 

\State \textbf{Step 3:}
Find $TM(c)$ \Comment{channel $c$'s optimal set of vehicles among those willing to transmit on $c$.} 
\State \indent Find $TM(v)$ \Comment{The set of channels preferred by $v$, among channels that the base station is willing to allocate to $v$.} 
\State Update $M(c) \leftarrow TM(c)$ and $M(v) \leftarrow TM(v)$
\State Iterate until two iterations give identical results. When two iterations are identical, stop.

\State \textbf{Step 4}
\State \indent \textbf{Return} $M(c),~M(v)$ \Comment{Stable NOMA resource allocation}
\end{algorithmic}
\end{algorithm}

To clarify Algorithm~\ref{algmatch}, we consider the following example.

\textbf{Example:} Consider a NOMA user-channel allocation with users $v_1, v_2,v_3, v_4, v_5$ and channels $c_1,c_2,c_3$. The preferences of users are in descending order. For instance, $P(v_1)= c_1c_2c_3, c_1c_2, c_1c_3, c_2c_3, c_1, c_2, c_3$ means user $v_1$ gains better QoS by transmitting over all three channels, which is preferred to both $c_1$ and $c_2$, preferred to both $c_1$ and $c_3$ and so on. Obviously, different users and channels have different preference orderings.



Step 1: First, we initialize the algorithm by allocating all the channels to every vehicle and allocating no car to any channel.

Step 2: For every vehicle $v$ and every channel $c$, using their given preferences and the matching $M$, the sets $U(c,M)$ and $V(v,M)$ happen to be as follows:
\begin{align}
\label{eq:step2}
&U(c_1,M) = U(c_2,M) = U(c_3,M)= v_1v_2v_3v_4v_5 \\ \nonumber
&V(v_1,M) = V(v_2,M) = V(v_3,M) = V(v_4,M) = V(v_5,M) = c_1c_2c_3
\end{align}
Step 3: For every vehicle $v$ and every channel $c$, using the matching $M$, find the sets $TM(v)$ and $TM(c)$ and update the current matching $M(.)$ using the value of $TM(.)$, i.e.,
\begin{align}
\label{eq:TM}
&TM(v_1) = c_1c_2c_3 \rightarrow M(v_1)\\ \nonumber
&TM(v_2)= c_1c_2c_3 \rightarrow M(v_2)\\ \nonumber
&TM(v_3)= c_1c_2c_3 \rightarrow M(v_3)\\ \nonumber
&TM(v_4)= c_1c_2c_3 \rightarrow M(v_4)\\ \nonumber
&TM(v_5) = c_1c_2c_3 \rightarrow M(v_5) \\ \nonumber
&TM(c_1) = v_1v_2 \rightarrow M(c_1)\\ \nonumber
&TM(c_2) = v_2v_3 \rightarrow M(c_2)\\ \nonumber
&TM(c_3) = v_3v_4 \rightarrow M(c_3)
\end{align}

In the second iteration of the algorithm, use the new values of matching $M$ obtained at the end of the previous iteration and restart from Step 2 in Algorithm~\ref{algmatch} to obtain new values of $U(c,M)$ and $V(c,M)$ as in Eq.~\ref{eq:iter2}.
\begin{align}
\label{eq:iter2}
&U(c_1,M) =\{v_1,v_2,v_3,v_4,v_5  \}\\ \nonumber
&U(c_2,M) = \{v_1,v_2,v_3,v_4,v_5 \}\\ \nonumber
&U(c_3,M) = \{v_1,v_2,v_3,v_4,v_5 \}\\ \nonumber
&V(v_1,M)=c_1\\ \nonumber
&V(v_2,M) = c_1c_2\\\nonumber
&V(v_3,M) = c_3\\ \nonumber
&V(v_4,M) = c_3\\ \nonumber
&V(v_5,M)  = \emptyset
\end{align}

Next, we move to Step 3 to update the values of NOMA resource allocation $M(c)$ and $M(v)$ for every channel $c$ and every vehicle $v$, as in Eq.~\ref{eq:iter2step3}
\begin{align}
\label{eq:iter2step3}
&TM(v_1) = c_1 \rightarrow M(v_1) \\ \nonumber
&TM(v_2) = c_1c_2 \rightarrow M(v_2)\\ \nonumber
&TM(v_3) =  c_3 \rightarrow M(v_3) \\\nonumber
&TM(v_4)  = c_3 \rightarrow M(v_4) \\\nonumber
&TM(v_5) = \emptyset \rightarrow M(v_5)\\\nonumber
&TM(c_1)  = v_1v_2 \rightarrow M(c_1)\\\nonumber
&TM(c_2)  = v_2v_3 \rightarrow M(c_2) \\\nonumber
&TM(c_3)  = v_3v_4 \rightarrow M(c_3) 
\end{align}

In the third iteration, we restart from Step 2 of Algorithm~\ref{algmatch} as in Eq.~\ref{eq:itr3}.
\begin{align}
\label{eq:itr3}
&U(c_1,M) = v_1v_2v_3v_4v_5\\\nonumber
&U(c_2,M)= v_1v_2v_3v_4v_5\\ \nonumber
&U(c_3,M) = v_1v_2v_3v_4v_5\\\nonumber
&V(v_1,M) = c_1 \\\nonumber
&V(v_2,M) = c_1c_2\\\nonumber
&V(v_3,M) = c_2c_3\\\nonumber
&V(v_4,M) = c_3 \\\nonumber
&V(v_5,M) = \emptyset
\end{align}
The matching $M(.)$ is updated according to Step 3 of Algorithm~\ref{algmatch} as follows: 
\begin{align}
&TM(v_1) =  c_1 \rightarrow M(v_1)\\ \nonumber
&TM(v_2) = c_1c_2 \rightarrow M(v_2)\\ \nonumber
&TM(v_3) = c_2c_3 \rightarrow M(v_3) \\\nonumber
&TM(v_4) = c_3 \rightarrow M(v_4) \\\nonumber
&TM(v_5) = \emptyset \rightarrow M(v_5)\\\nonumber
&TM(c_1) = v_1 v_2 \rightarrow M(c_1) \\\nonumber
&TM(c_2) = v_2v_3\rightarrow M(c_2) \\\nonumber
&TM(c_3) = v_3v_4 \rightarrow M(c_3) 
\end{align}
In the fourth iteration, the same results are obtained as in the third iteration. It shows that the algorithm has converged to a stable solution.  

\subsection{Stability Analysis and Convergence of Algorithm~\ref{algmatch}}

The gist of NOMA resource allocation in Algorithm~\ref{algmatch} is finding the fix-points of a function $TM(.)$ using the matching $M$~\cite{talgorithm}. The fix-points of a function are values in the domain of the function that are mapped to the same equal value (i.e., to themselves) in the range. Algorithm~\ref{algmatch} finds the fix-points by iteration. As soon as two iterations are the same, the fix-point has been found, and the result is a stable matching.

Even for non-substitutable preferences of users and channels, the matching returned by Algorithm~\ref{algmatch} is always pairwise stable~\cite{oveido}. In other words, this NOMA resource allocation is robust even if neither the vehicles' preferences $p(v)$,~$\forall v\in \mathscr{V}$ nor the channels' preferences $p(c)$,~$\forall c\in \mathscr{C}$ meet the substitutability property.

In addition, if only the preferences of the vehicles $p(v),~\forall v\in \mathscr{V}$ are substitutable, while the preferences of the channels  $p(c),~\forall c\in \mathscr{C}$ are not substitutable, the NOMA resource allocation returned by Algorithm~\ref{algmatch} is stable and coincides with the subgame perfect Nash equilibrium of the game described in Section III.~B and Fig.\ref{fig:spne}. By symmetry, the same result holds~\cite{oveido} if only the preferences of the channels $p(c),~\forall c\in \mathscr{C}$ are substitutable, while the preferences of the vehicles $p(v),~\forall v\in \mathscr{V}$ do not satisfy the substitutability property.  

Asymptotically, in the limits, if one side, i.e., $p(c),~\forall c\in \mathscr{C}$ is substitutable and the other side, e.g., $p(v),~\forall v\in \mathscr{V}$ is strongly substitutable, the resource allocation returned by Algorithm~\ref{algmatch} is always guaranteed to be setwise stable. For example, this might happen in practice when a quota is imposed on the maximum number of NOMA users per channel to restrict the complexity of the SIC decoder at the receiver. In this case, strong substitutability may be challenging to implement or not feasible for $p(c),~\forall c\in \mathscr{C}$. However, with the advent of advanced NOMA decoders that can handle massive number of users per channel, a setwise stable NOMA resource allocation is always guaranteed to exist.

To ease the above-mentioned condition on strong substitutability, if we let both sides, i.e., $p(c),~\forall c\in \mathscr{C}$ and $p(v),~\forall v\in \mathscr{V}$ be only substitutable, the NOMA user-channel allocation scheme described in Algorithm~\ref{algmatch} is guaranteed to converge to an answer, which is pairwise stable~\cite{oveido}. 

\section{Conclusion}
\label{sec:conclus}
This paper has aimed to investigate the problem of NOMA resource allocation for a set of users and a set of channels. Every channel can be allocated to more than one user by utilizing superposition coding. Each user can operate on multiple channels simultaneously as long as this is allowed by the hardware/software radio transceiver configurations of the user. The proposed resource allocation was studied for cellular V2X. The base station holds the vehicles' geolocation information. The conditions for the stability of resource allocation were further discussed in this paper. Additionally, a stable NOMA user-channel allocation algorithm was used that is individually rational.  Unlike OFDMA, the stable NOMA user-channel allocation algorithm circumvents nonoptimality of core allocations.    

\bibliography{IEEEabrv,arxiv}
\bibliographystyle{ieeetran}

\end{document}